\documentstyle[11pt,newpasp,twoside]{article}
\def\edcomment#1{\iffalse\marginpar{\raggedright\sl#1\/}\else\relax\fi}
\reversemarginpar

\begin{document}
\title{The spectroscopic orbits and other parameters of the symbiotic binary FN Sgr}

\author{E. Brandi$^{1,2}$, J. Miko{\l}ajewska$^{3}$, C. Quiroga$^{1,4}$, K. Kulczycki$^{3}$,
K. Belczy{\'n}ski$^{3}$, O.E. Ferrer$^{1,4}$, L.G. Garc\'{\i}a$^{1}$ and C.B. 
Pereira$^{5}$ }

\affil{$^1$ Facultad de Ciencias Astron\'omicas y Geof\'{\i}sicas, Universidad
Nacional de La  Plata. Paseo del Bosque S/N - 1900 la Plata, Argentina}

\affil{$^2$ Comisi\'on de Investigaciones Cient\'{\i}ficas de la Provincia de 
Buenos Aires (CIC), Argentina} 

\affil{$^3$ Copernicus Astronomical Center, Warsaw, Poland} 

\affil{$^4$ Consejo Nacional de Investigaciones Cient\'{\i}ficas y T\'ecnicas
(CONICET), Argentina}

\affil{$^5$ Observatorio Nacional -- Rua Gen. Jos\'e Cristino, 77, Sao Cristovao -
CEP 20921-400, Rio de Janeiro, Brasil }

\begin{abstract}

We present an analysis of optical and near infrared spectra 
of the eclipsing 
symbiotic system FN Sgr. In particular,  we have determined for the first time 
spectroscopic orbits  based of the radial velocity curves for both  components. A set of 
blue absorption lines resembling an A-F type star is present in all our spectra. They seem 
to be associated with the hot component but the best orbital fitting corresponds to a 
larger eccentricity.  We have also studied spectral  changes and photometric
variations in  function of both  the hot component activity and the orbital
motion. 
\end{abstract}

\section{Introduction}

This work is part of a major project in which we determine periods and
spectroscopic orbits of symbiotic binaries in base of observational data
collected at CASLEO (San Juan, Argentina) along more than ten years.     
Selecting the southern S-type symbiotic star FN Sgr, we chose to use visual  
photometry from the Variable Star Section Circulars of The
Royal Astronomical Society of  New Zealand through 20 years of these circulars
(1975$-$2001; Fig.1) and high  resolution spectra obtained with the 2.15 m
telescope of CASLEO during the period 1990$-$2001. 

We have measured the radial velocities of the cool component from the M-type absorption 
lines, specially FeI, TiI, NiI, SiI and CoI. We have also identified and measured the blue 
cF-type absorption lines corresponding to CrII, FeII, TiII and YII which are 
believed to be linked to the hot companion (e.g. Miko{\l}ajewska \& 
Kenyon 1992; Quiroga et al. 2002). The individual radial velocities were obtained by 
gaussian fitting of the line profiles an a mean value was calculated for each spectrum.  
In addition, we have determined the radial velocities of the broad emission wings of 
H$\alpha$, H$\beta$ and HeII$\lambda$4686 which are formed in the inner region of the 
accretion disk or in an extended envelope near the hot component (e.g. Quiroga et al. 
2002). For this we have used the method outlined by Schneider \& Young (1980).

Optical emission line fluxes were also derived,  either by integrating the line profile or 
by fitting Gaussian profiles, as well as the [TiO]$_1$ and [TiO]$_2$ indices as 
defined by Kenyon \& Fernandez-Castro (1987).

\begin{figure}
\plotfiddle{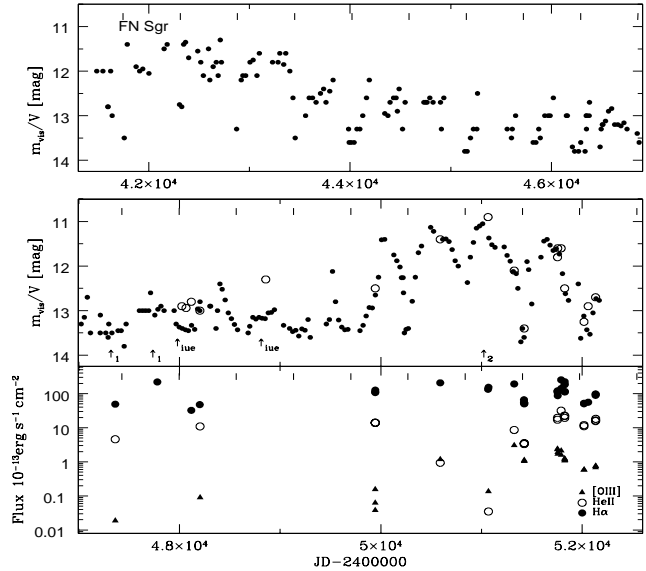}{2.55in}{0deg}{45}{40}{-140in}{-75in}
\caption{(top)Visual light curve of FN Sgr in 1975$-$2002. Dots:
visual observations from RASNZ, open circles:  V magnitudes calculated from
our spectra and  derived from FES counts. (bottom) Evolution  of emission line
fluxes of H$\alpha$, HeII\,4686 and [OIII\,5007] in 1988$-$2002. 
Bars mark times of photometric minima/times of inferior spectroscopic
conjunction of the red giant.}  
\end{figure} 

\begin{figure} 
\plotfiddle{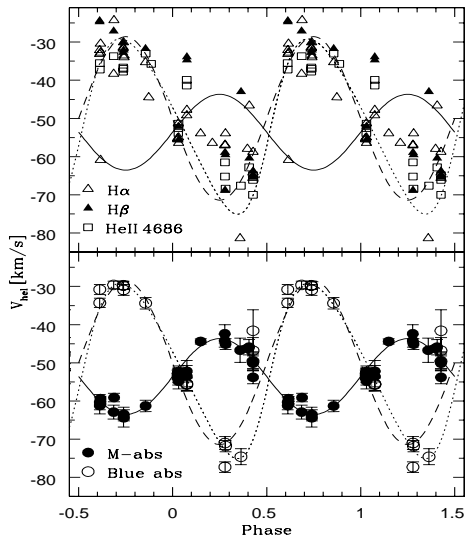}{2.5in}{0deg}{50}{40.5}{-140in}{-90in}
\caption{Phased radial velocity data and orbital solutions for FN Sgr. (top)
The broad emission wings of  HI and HeII lines. (bottom) The M giant  (dots)
and the hot component (open circles). The solid and dashed lines repeat the
circular orbit of the M giant and the hot component  (blue absorption system),
respectively. The dotted line gives the best elliptical fit to the blue
absorption system ($e=0.20$).} 
\end{figure}

\section{Orbital period}

 We have analyzed the RASNZ visual photometry using the period-search method 
described by Schwarzenberg-Czerny (1997). A period of about 568 days was
obtained from all visual data presented in Fig.1. The radial
velocities, emission line fluxes, and [TiO]$_1$ indices,
all  show the same periodicity, which we attribute to the orbital period. In
all cases, the  most regular changes have been obtained with 568-day period,
giving the ephemeris  \begin{equation}  \mbox{Min} = \mbox{JD}\,2450270\,(\pm
2) + 568.3\,(\pm 0.3)\, \times \mbox{E}.  \label{aee} 
\end{equation}

\section{Spectroscopic orbits}

 The broad emission line wings of H$\alpha$, H$\beta$ and HeII\,4686 and the blue cF 
type absorption lines show the highest amplitude and a mean velocity similar
to the red giant  systemic velocity. They are clearly in antiphase with the
M-giant curve which suggest that  they are fomed in a same region very near to
the hot component (Fig.2). 

\begin{table}
\caption{Orbital solutions for FN Sgr} 
\begin{tabular}{lccccc} 
\tableline
Elements & \multicolumn{2}{c}{M giant} & \multicolumn{2}{c}{Blue
absorptions} & HeII wings \\
\tableline
$K$    [km/s]       & $10.5\pm0.4$  & $10.6\pm0.3$  & $21.7\pm1.0$
&$22.9\pm0.6$ &$17.4\pm0.7$ \\   $\gamma_0$ [km/s] &
$-53.7\pm0.2$&$-53.7\pm0.2$ & $-50.9\pm0.9$&$-51.1\pm0.7$ &  $-50.5\pm0.7$\\ 
$e$       & 0            & $0.07\pm0.03$ & 0       &  $0.20\pm0.03$&
$0.28\pm0.04$ \\  $\omega$  [\deg]    &              & $320\pm26$    &    
&$257\pm8$ & $261\pm13$ \\  $T_{0}$ [JD\,24...] &               & 50355       
&      & 50541 & 50561 \\  $\Delta T$ [day]    &   7.4         & 13           
& 3.4           &  11 &  \\  $a \sin i$ [R$_{\sun}$] & 118  & 118           &
243           & 252 & 187 \\  $f(M)$ [M$_{\sun}$] & 0.0689    & 0.0694       
& 0.5996        & 0.6670 & 0.2722 \\     \tableline
\tableline
\end{tabular} 
 $T_{0}$  -- time of the passage through periastron;~~ 
 $\Delta T = T_{\rm sp\,conj}-T_{\rm phot\, min}$.

\end{table}

The best orbital solution for the blue absorption system (as well as those for the broad 
emission line wings) leads to significant eccentricity whereas the red giant orbit is 
circular or nearly circular.

Combining the semi-amplitudes of the M giant and the blue absorption component for the 
circular orbit (Table 1) gives a mass ratio $q=M_{\rm g}/M_{\rm h}=2.1 \pm
0.2$,  the  component masses of $M_{\rm g} \sin^3 i = 1.4\, \rm M_{\sun}$ and
$M_{\rm h} \sin^3 i =  0.66\, \rm M_{\sun}$, and the binary separation $a \sin
i = 360\, \rm R_{\sun}$.

\section{Activity}

 Most of our spectroscopic observations were taken during the optical outburst phase 
(Fig.1). The outburst behaviour is similar to that observed in other classical 
symbiotic stars (e.g. CI Cyg and AX Per). The rise in optical brightness was first 
accompanied by a large increase in the [OIII] emission lines, and broadening of the 
emission line wings.  The permitted HI and HeI emission lines 
were also increasing although not as much as the forbidden lines, whereas HeII\,4686 was 
decreasing.
Then, near the visual maximum (1996$-$1998) the [OIII] emission decreased, and HeII\,4686
practically disappeared. At the same time the permitted emission lines have maximum 
widths. In 1999,  the visual brightness began gradually declining; the [OIII] and 
HeII emission 
lines were increasing again. The [OIII] nebular emission reached maximum intensity in 
July$-$September 2000, and after that was declining following the visual magnitudes.
HeII\,4686 emission reached maximum intensity in September$-$October 2000, somewhat later
than the forbidden line emission.\\
The emission line changes indicate a decrease in the hot component temperature at the 
optical maximum to $\la 50\,000\, \rm K$, possibly because of increasing
optical depth in  the hot component wind. On the other hand, the presence of
the blue shell absorption  system apparently associated with the hot component
in all our spectra, including periods  where the strong high-ionization
features are present, suggests a double-temperature  structure of the hot
component. The emission line fluxes, if due to a photoionization, require
the hot component temperature and luminosity of $T_{\rm h} \sim 150\,000\, \rm
K$,  $L_{\rm h} \sim 1000\, \rm L_{\sun}$  at quiescence, and  $T_{\rm h} \sim
50\,000\, \rm  K$,  $L_{\rm h} \sim 2000\, \rm L_{\sun}$ at the optical
maximum. The optical observed  during the 1996-98 maximum is consistent with a
presence of continuum source with $T_{\rm  eff} \sim 7000 - 8000\, \rm K$ and
$L \sim 2000 - 1000\, \rm L_{\sun}$ (assuming a  distance $d=5\, \rm kpc$). 

We note here that similar behavior was observed during the optical outbursts of 
AX Per (Miko{\l}ajewska \& Kenyon 1992), and other active symbiotic systems,
in  particular, in AR Pav (Quiroga et al. 2002, and references therein), and
it can be explained by a presence of optically thick accretion disks.

\section{ Acknowledgments}
 This research was partly founded by KBN Research Grant No. 5\,P03D\,019\,20.


\begin{references}
\reference Kenyon, S.J., \& Fernandez-Castro, T. 1987, \aj, 93, 938 
\reference Miko{\l}ajewska, J., \& Kenyon, S.J. 1992, \aj, 103, 579 
\reference Quiroga, C., Miko{\l}ajewska, J., Brandi, E., Ferrer, O., \& 
Garc\'{\i}a, L. 2002, \aap, 387, 139   
\reference Schneider, D. P., \& Young, P. 1980, \apj, 238, 946 
\reference Schwarzenberg-Czerny, A. 1997, \apj, 489, 941

\end{references}
\end{document}